\documentclass[epj,nopacs]{svjour}
\usepackage{graphicx}
\usepackage{xspace}
\usepackage{amsmath}
\usepackage{url}

\newcommand{\DZero}{D\O\xspace}

\newcommand{\cer}{\v{C}erenkov\xspace}

\newcommand{\cp}{\emph{CP}\xspace}

\newcommand{\mathsl}[1]{\mbox{\textsl{#1}}}

\newcommand{\meson}[1]{\ensuremath{\mathsl{#1}}}
\newcommand{\baryon}[1]{\ensuremath{\mathsl{#1}}}
\newcommand{\antibaryon}[1]{\ensuremath{\overline{\mathsl{#1}}}}
\newcommand{\quark}[1]{\ensuremath{\mathsl{#1}}}
\newcommand{\antiquark}[1]{\ensuremath{\overline{\mathsl{#1}}}}
\newcommand{\lepton}[1]{\ensuremath{\mathsl{#1}}}

\newcommand{\cq}{\quark{c}\xspace}
\newcommand{\cbq}{\antiquark{c}\xspace}
\newcommand{\bq}{\quark{b}\xspace}
\newcommand{\bbq}{\antiquark{b}\xspace}

\newcommand{\electronneg}{\lepton{e}^-\xspace}
\newcommand{\positron}{\lepton{e}^+\xspace}
\newcommand{\muon}{\ensuremath{\mu}\xspace}

\newcommand{\proton}{\baryon{p}\xspace}
\newcommand{\pbar}{\antibaryon{p}\xspace}

\newcommand{\pion}{\ensuremath{\pi}\xspace}

\newcommand{\pizero}{\ensuremath{\pion^0}\xspace}

\newcommand{\kaon}{\ensuremath{\meson{K}}\xspace}

\newcommand{\jpsi}{\ensuremath{J\!/\!\psi}\xspace}

\newcommand{\bmeson}{\ensuremath{\meson{B}}\xspace}
\newcommand{\bbar}{\ensuremath{\bar{\meson{B}}}\xspace}
\newcommand{\bplus}{\ensuremath{\bmeson^{+}}\xspace}

\newcommand{\bzero}{\ensuremath{\bmeson{}^{0}}\xspace}
\newcommand{\bszero}{\ensuremath{\bmeson{}^{0}_s}\xspace}

\newcommand{\gevc}{\ensuremath{\mathrm{GeV}/c}\xspace}

\newcommand{\micron}{\ensuremath{\mu\mathrm{m}}\xspace}

\newcommand{\degree}{\ensuremath{^\circ}\xspace}

\newcommand{\figref}[1]{Figure~\ref{fig:#1}}

\begin{document}
\title{The BTeV Experiment}
\author{Eric W. Vaandering\inst{1} 
\thanks{on behalf of the BTeV Collaboration}
}                     % Do not remove
\institute{Vanderbilt University, Nashville, TN 37013, U.S.A.}
\date{Received: \today / Revised version: \today}
\abstract{BTeV is an approved forward collider experiment at the Fermilab
Tevatron dedicated to precision studies of CP violation, mixing, and rare
decays of beauty and charmed hadrons. The BTeV detector has been designed to
achieve these goals. Pixel detectors cover the interaction region and vertex
computation is included in the lowest level trigger.
\PACS{
      {PACS-key}{discribing text of that key}   \and
      {PACS-key}{discribing text of that key}
     } % end of PACS codes
} %end of abstract
\maketitle
\section{Introduction}\label{sec:intro}

The BTeV experiment~\cite{www_btev_natbib} is an approved forward collider experiment at the Fermilab
Tevatron dedicated to \bmeson physics. The experiment will begin data taking in
2009. 

\begin{figure}
\begin{center}
\includegraphics[bb=25 160 570 640,width=8.5cm,clip]{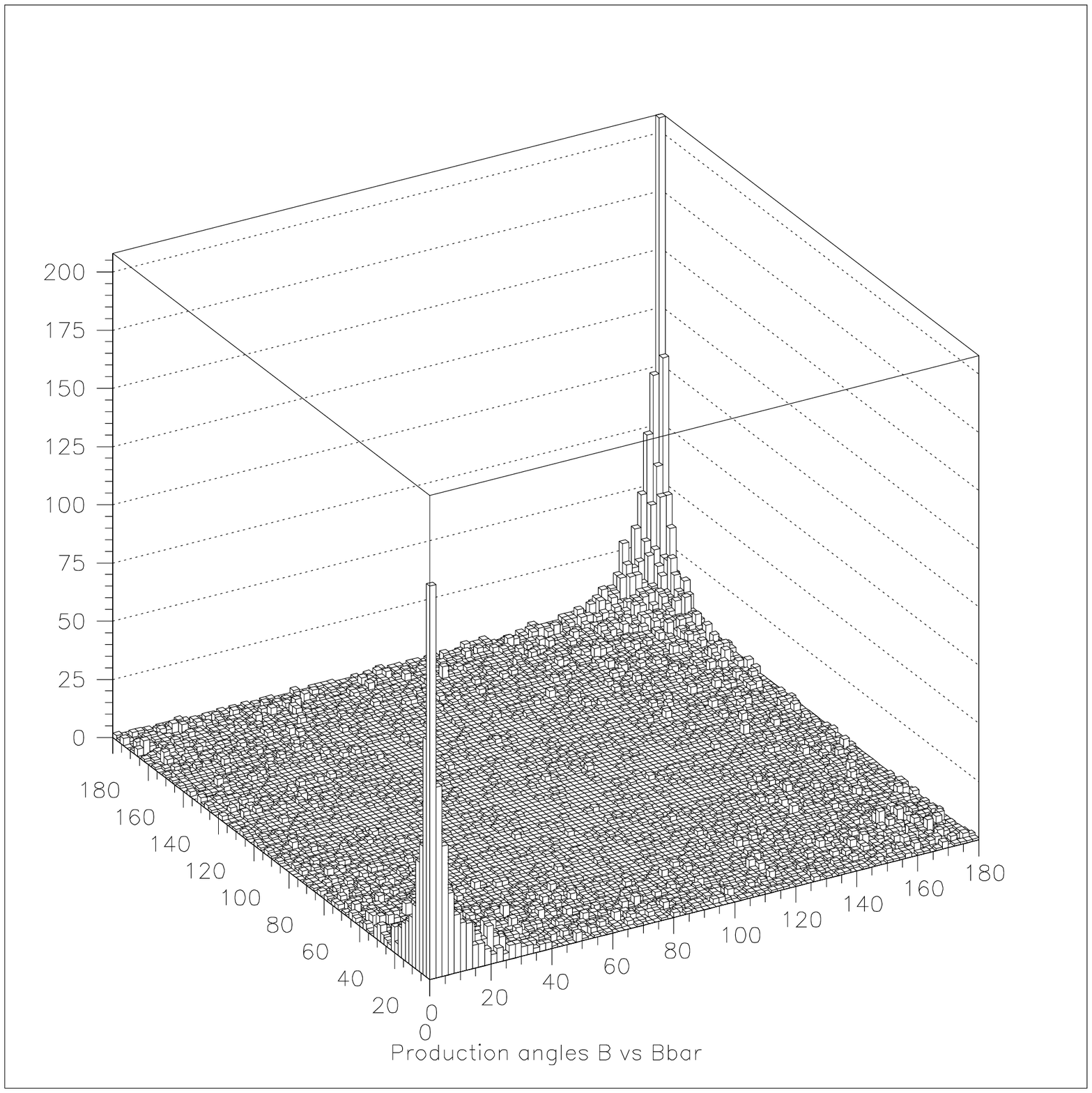}
\end{center}
\caption{Production angles for \bmeson and \bbar a the Tevatron. Large
numbers of correlated  $\bmeson\bbar$ pairs are produced at 0\degree and
180\degree.}
\label{fig:forward_b}       % Give a unique label
\end{figure}

Measurements by \DZero have shown that at the Tevatron the $\bq\bbq$ cross
section is at least 100~$\mu$b~\cite{dzero_for_xs}; the $\cq\cbq$ cross section is expected to
be an order of magnitude higher. Much of this production is in the forward
direction. Furthermore, as shown in \figref{forward_b}, when the \bq hadron is in the forward direction, the
\bbq hadron is also likely to be in the forward direction. This is especially
important for mixing and \cp analyses where flavor tagging is crucial. 

The fraction of interactions which produce $\bmeson\bbar$ pairs is
approximately 0.2\%; BTeV is expected to have a \bmeson triggering efficiency
of $>50$\%. Running at an average luminosity of
$2\times10^{32}$~cm$^{-2}$s$^{-1}$, BTeV will reconstruct about
$2\times10^{11}$ $\bmeson\bbar$ pairs/year. 

\section{B Physics at Hadron Colliders}

Despite (and in addition to) these high rates of \bmeson production, \bmeson
physics at a hadron collider presents additional opportunities and challenges
when compared to $\positron\electronneg$ collider experiments.

One of the primary advantages provided by hadron colliders is the production of
all species of \bq hadrons, not just \bzero and \bplus. This is important,
because to test the consistency of the Standard Model description of \cp violation, one
must also study \cp violation in the \bszero sector. The \bszero analogue to
$\beta$ in \bzero mixing, $\chi$, is expected to be small (a few degrees).
Additionally, not much information is known about the $B_c^+$, $\Lambda_b^0$, 
$\Xi_b$, and $\Omega_b$ states. Results from hadron machines will be crucial in
understanding the \bq baryons. 

Another advantage of the forward hadron collider environment
is that relativistic boosts of the particles are large, making it
comparatively easy to measure particle lifetimes with great accuracy. This is
especially useful in mixing measurements; while $\Delta\Gamma$ in \bszero
mixing may be measured at CDF and \DZero, mixing angle measurements of
the \bszero require this precise lifetime measurement capability.

\bmeson physics in a hadron environment is also challenging. The lack
of $4\pi$ coverage removes the ability to use beam constraints. Additionally,
the larger fraction of background events and the larger multiplicities makes
electromagnetic calorimetry and particle ID more challenging.  Moving
the study of \cp violation beyond $\sin 2\beta$ to redundant measurements of
the other \cp violating angles $\alpha$, $\beta$, $\gamma$, and $\chi$ requires effective
photon detection.

\bmeson factories at hadron colliders are well positioned to make these crucial
measurements.

\section{The BTeV Spectrometer}\label{sec:spect}

\begin{figure*}
\begin{center}
  \includegraphics[width=0.75\textwidth]{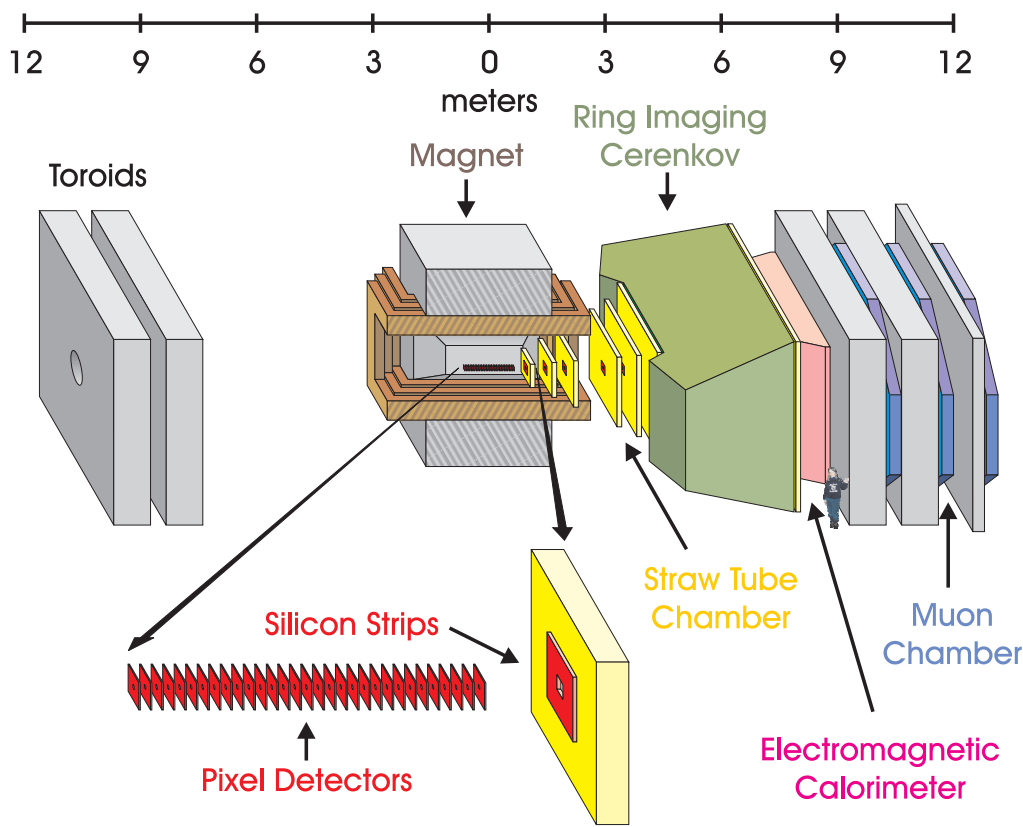}
\end{center}
\caption{The BTeV Spectrometer}
\label{fig:spect}       % Give a unique label
\end{figure*}

Because it instruments the forward region (10--300~mrad), the BTeV spectrometer resembles past
fixed target experiments in overall layout. \figref{spect} shows a schematic
overview of the BTeV spectrometer. $\proton\pbar$ interactions occur at the
center of a large dipole magnet. These interactions are traced, within the
magnetic field, by a silicon pixel detector. ``Downstream'' tracking is
performed with silicon trackers in the most forward regions and straw tube
trackers in the lower occupancy regions. A Ring Imaging \cer (RICH) is used
for charged particle identification, electrons and photons are reconstructed in
a lead tungstate (PbWO$_4$) crystal calorimeter, and muon identification is
performed with muon detectors shielded by magnetized steel toroids.

\textbf{Pixel Detector:}
The pixel detector sits inside a 1.6~T magnetic field. By providing accurate 3-D
space points, it makes the challenging vertex trigger (described later) possible
as well as providing accurate vertexing information for physics reconstruction.
The detector consists of 30 stations of doublets along the beam direction with a
pixel size of $50\times400$~$\mu$m. There are approximately 23 million channels
in the full detector.

\textbf{Tracking Detectors:} Additional tracking in BTeV is provided by
seven stations of tracking chambers, each of which consists of silicon strip
detectors in the forward region and straw detectors in the outer region. Three
of these stations are located within the magnetic field, the rest are outside
it. The silicon detectors have a pitch of 100~\micron and cover the central
$27\times27$~cm$^2$ at each station. The straws have a diameter of 4~mm and are
arranged in 3 views, each of which has three layers arranged in a close packed
geometry.

\textbf{RICH Detector:} For charged particle identification, BTeV will use a
RICH detector. In order to obtain good separation between particle species over
a range of momentum, two radiators will be used: a liquid radiator (a thin
layer of C$_5$F$_{12}$ at the front of the detector) and a large gas radiator
volume of C$_4$F$_{10}$. Photons from the gas radiator will be reflected and
detected on 163-channel Hybrid Photodiodes (HPD) or 64-channel Multi-Anode
Photomultiplier Tubes (MAPMT) while photons from the liquid
radiator will be directly detected with 3'' PMTs which,
due to the large \cer angles will be mounted on the sides of the detector. The
combination of these two radiators will provide good \kaon-\pion and \proton-\kaon
separation out to 70~\gevc. Additionally, in the wide angle, low momentum region not
covered by the muon detector (as described below) the RICH will be able to
provide \pion-\muon separation.

\textbf{Electromagnetic Calorimeter:}
Photon and electron reconstruction will use a PbWO$_4$
(lead tungstate) crystal calorimeter. These are the same crystals being used by CMS. Each
crystal will be $2.8\times2.8\times22$~cm$^3$. Because these crystals are very
fragile, each crystal (or at most, small groups of crystals) must be
individually supported. Lead tungstate provides excellent energy resolution; in
beam tests, BTeV has found an energy resolution of 
\begin{equation*}
\sigma_E/E = (0.33 + 1.8 / \sqrt{E})\% \, .
\end{equation*}
Combined with the small crystal size, this will give BTeV electromagnetic
calorimetry similar in performance to detectors at $\positron\electronneg$
machines.

\textbf{Muon Detector:}
Muon identification in BTeV will be provided by a system of proportional
counters. There are three stations of muon detectors with four views per station
(2 $r$ views, $u$, and $v$ views). Each view consists of layers of stainless
steel tubes (3/8'' diameter) arranged in a picket fence geometry. ``Upstream''
of the first station and between the first and second stations are 1~m thick
magnetized steel toroids which absorb other particles and also allow a stand-alone
momentum measurement.

The momentum measurement also allows a completely independent \jpsi trigger at the first
level to enhance samples of interesting decays and to provide an unbiased selection
of events for calibrating the crucial first level vertex trigger. 

\section{Detached Vertex Trigger}\label{sec:trigger}

To collect the maximum number of \bmeson decays, BTeV will employ a detached
vertex trigger at the first level which will partially reconstruct every event.
Every beam crossing is read out by the DAQ and stored in 1~TB of buffer memory.
Crossings will occur in the Tevatron with a maximum rate of 7.6~MHz if
132~ns bunch crossings can be achieved (396~ns is the current expectation). 

First level triggering will be performed by a cluster of Field Programmable Gate
Arrays (FPGA) and about 2500 Digital
Signal Processors (DSP) for the vertex trigger and an additional cluster of
about 500 DSPs for the di-muon trigger. This ambitious vertex trigger is only
possible because of the 3D space points provided by the pixel detector. 

The detached vertex trigger algorithm begins with FPGAs assembling vertex hits
into track segments. Hits near the interaction region and near the edges of the
pixel planes are used. The inner region hits help define intercepts and vertex
positions, the outer hits are used to refine the track vectors and to measure
the momenta as the particles are bent in the magnet.

These track segments are then passed to the farm of DSPs for segment matching
and vertex reconstruction. The DSP-based trigger algorithm looks for a
production vertex and tracks that miss that vertex. Our current algorithm
requires two tracks that miss the production vertex by $>6\sigma$. Because a
momentum measurement is also performed, we can also put requirements on $p_T$
of the tracks. This will be essential in removing the more copious charm
decays, which are largely a background for BTeV.

%\begin{table}
%\begin{tabular}{lll}
%\hline\noalign{\smallskip}
%\end{tabular}
%\end{table}
%

\section{Comparison with LHCb}

BTeV's primary competition will come from the LHCb experiment at CERN. Both will
begin data taking at about the same time and aim to do much of the same physics,
however there are significant differences in the detector designs.

LHCb has several advantages over BTeV:
\begin{itemize}
\item{The $\bq\bbq$ cross section at the LHC is expected to be about five times
higher than at the Tevatron while the total cross section is only about 1.6
times higher.}
\item{LHCb will operate with less than one interaction per beam
crossing, while BTeV will operate with 2--6 depending on the crossing time.
However, the crossing time at LHC is 25~ns, somewhat below the response times of
most detectors.}
\end{itemize}

BTeV will have a number of advantages over LHCb as well:

\begin{itemize}
\item{BTeV has a dipole located on the interaction region which gives it a spectrometer covering the forward
antiproton rapidity region. This will aid in locating production vertices.}

\item{BTeV uses a precision vertex detector based on planar pixel arrays.}

\item{The pixel detector enables a vertex trigger at Level 1. This makes BTeV especially efficient for states
that have only hadrons and allows for less restrictive definitions of
``interesting'' events.}

\item{The pixel detector also helps minimize confusion for multiple interactions per
crossing.}

\item{BTeV has a lead tungstate electromagnetic calorimeter, giving it very good
capabilities for photon and \pizero
reconstruction.}

\item{BTeV plans a very high capacity data acquisition system which frees it from making
excessively restrictive choices at the trigger level. This will give us an
unbiased selection of \bq and \cq decays and means that as new physics becomes
interesting, we will have those events ``on tape'' already.}
\end{itemize}

\section{Conclusions}

The frontier in studies in \bmeson flavor physics will move to hadron colliders.
The BTeV experiment, with its innovative design and triggering, will help make
the next decade in \bmeson physics as interesting as the current one.

% BibTeX users please use

\bibliographystyle{phaip}
\bibliography{btev}

\end{document}